\documentclass[10pt,preprint2,longabstract]{aastex}

\newcommand{\tsco}{$\tau$\,Sco}

\usepackage{txfonts}
\usepackage{epsf}
\usepackage{graphicx}
\usepackage{graphics}
\usepackage{color}
\usepackage{subfigure}
\usepackage{psfig}

\def \etal   {\hbox{et~al.\/}}

\newcommand{\bvec}[1]{\mbox{\boldmath ${#1}$}}

\shorttitle{X-rays from $\tau$ Sco}
\shortauthors{Ignace et al.}

\begin{document}

\title{A Multi-Phase {\it Suzaku} Study of X-rays from $\tau$ Sco}

\author{R.~Ignace}
\affil{Department of Physics and Astronomy, East Tennessee State University,
Johnson City, TN 37614, USA}
\email{ignace@etsu.edu}

\author{L.~M.~Oskinova}
\affil{Institute for Physics and Astronomy, University Potsdam, 
14476 Potsdam, Germany}
\email{lida@astro.physik.uni-potsdam.de}

\author{M.~Jardine}
\affil{School of Physics and Astronomy, University of St~Andrews,
St~Andrews, Scotland KY16 9SS, UK}
\email{moira.jardine@st-and.ac.uk}

\author{J.~P.~Cassinelli}
\affil{Department of Astronomy, University of Wisconsin,
Madison, WI 53706 USA}
\email{cassinelli@astro.wisc.edu}

\author{D.~H.~Cohen}
\affil{Department of Physics \& Astronomy, Swarthmore College,
Swarthmore, PA 19081, USA }
\email{cohen@astro.swarthmore.edu}

\author{J.-F.~Donati}
\affil{LATT, Observatoire Midi-Pyrenees, 14 Av.\ E.\ Belin, F-31400
Toulouse, France }
\email{donati@ast.obs-mip.fr}

\author{R.~Townsend}
\affil{Department of Astronomy, University of Wisconsin,
Madison, WI 53706, USA}
\email{townsend@astro.wisc.edu}

\author{A.~ud-Doula}
\affil{Department of Physics, Penn State Worthington Scranton, 
120 Ridge View Dr, Dunmore, PA 18512, USA}
\email{uddoula@gmail.com}



\begin{abstract} 

We obtained relatively high signal-to-noise X-ray spectral data of
the early massive star $\tau$~Sco (B0.2V) with the {\it Suzaku} XIS
instrument.  This source displays several unusual features that motivated
our study:  (a) redshifted absorption in UV P~Cygni lines to approximately
$+250$ km~s$^{-1}$ suggestive of infalling gas, (b) unusually hard X-ray
emission requiring hot plasma at temperatures in excess of 10~MK whereas
most massive stars show relatively soft X-rays at a few~MK, and (c)
a complex photospheric magnetic field of open and closed field lines.
In an attempt to understand the hard component better, X-ray data were
obtained at six roughly equally spaced phases within the same epoch
of $\tau$~Sco's 41~day rotation period.  The XIS instrument has three
operable detectors:  XIS1 is back illuminated with sensitivity down to
0.2~keV; XIS0 and XIS2 are front illuminated with sensivitity only down
to 0.4~keV and have overall less effective area than XIS1.  The XIS0
and XIS3 detectors show relatively little variability.  In contrast,
there is a $\approx 4\sigma$ detection of a $\approx 4\%$ drop in the
count rate of the XIS1 detector at one rotational phase.  In addition,
all three detectors show a $\approx 3\%$ increase in count rate at the
same phase.  The most optimistic prediction of X-ray variability allows
for a 40\% change in the count rate, particularly near phases where we
have pointings.  Observed modulations in the X-ray light curve on the
rotation cycle is an order of magnitude smaller than this, which
places new stringent constraints on future modeling of this interesting
magnetic massive star.

\end{abstract}


\maketitle

\section{Introduction}

The B0.5V star $\tau$~Sco has a storied history in the study of
massive stars.  A northern hemisphere object of early spectral type
that is relatively bright and close has made $\tau$~Sco a favored
target for multi-wavelength studies.  It was a primary target in
the early days of ultraviolet astronomy (e.g., Rogerson \& Lamers
1976; Rogerson \& Upson 1977) and was found to be a relatively hard
X-ray source as compared to its peers (MacFarlane \& Cassinelli
1989; Cassinelli \etal\ 1994).  It also displays redshifted absorption
in UV P~Cygni lines of O{\sc vi} and N{\sc v} (Lamers \& Rogerson
1978).  These features combined with growing evidence of clumped
wind flows motivated Howk \etal\ (2000) to develop a model of
infalling wind clumps to explain the observed redshifted absorptions
and as a way of producing high velocity shocked gas to account for
significant amounts of high temperature gas in excess of 10~MK in the
system.  X-ray data from {\it Chandra} (Cohen \etal\ 2003) and {\it
XMM-Newton} (Mewe \etal\ 2003) confirmed the unusual hardness of
this source, revealing also that the emission lines are relatively
narrow.  Since then, advances in observational spectropolarimetry
have allowed for more detections and more stringent constraints of
magnetism in massive stars (e.g., Donati \etal\ 2006a; Hubrig \etal\
2008; Alecian \etal\ 2008).  Donati \etal\ (2006b) reported on the
discovery of a surprisingly complex photospheric magnetic field in
$\tau$~Sco, a topology that sports sectors of open field lines and
confining loops that deviate significantly from a simple dipole
morphology.

All of these observational features together have yet to be explained by
a single unifying model.  The emission lines resolved by {\em Chandra}
span a large range of energies, and yet all of the lines are quite narrow,
with typical HWHM velocities of 300 km s$^{-1}$.  From Gaussian fits,
the peak emissions for an ensemble of lines show little net velocity
shift (Cohen \etal\ 2003).  Both of these features suggest minimal
photoabsorption by the wind of $\tau$~Sco (e.g., Owocki \& Cohen 2001).
It may also be that the wind is relatively porous.  (Discussions of the
impact of clumping and porosity for X-ray emissions and line profile
shapes can be found in Feldmeier, Oskinova, \& Hamann (2003), Oskinova,
Feldmeier, \& Hamann (2004), and Owocki \& Cohen (2006).

An understanding of the X-ray hardness and line profile properties is
critical for producing a quantitative physical model for $\tau$~Sco's
wind structure.  One way to further constrain the modeling effort is to
analyze variability of $\tau$~Sco.  Observing time of $\tau$~Sco using
{\it Suzaku} was awarded to obtain multiple pointings with rotational
phase to constrain the relative contributions of confined plasma to
wind shocked plasma in terms of the X-ray emission measure and plasma
temperature.  Our experiment was designed to take advantage of the
following facts.  First, $\tau$~Sco is a slow rotator with a rotation
period of approximately 41~days.  It is also relatively bright in X-rays,
and so exposures for the different pointings are quite short as compared
to the rotation period.  In addition, the star is viewed nearly edge-on
with respect to the rotational equator such that the closed magnetic
loops, which are asymmetrically distributed around the star, are occulted
during a rotation cycle.  In fact, Donati \etal\ (2006b) predicted
that the X-ray emissions of $\tau$~Sco could vary by as much as
40\% with rotational phase.  

Given that the field topology has remained stable over the last
decade (Donati \& Landstreet 2009), multiple pointings -- even if
at different epochs -- have the potential of determining hot plasma
components in the closed magnetic loops as compared to the larger
scale wind through an analysis of X-ray variability.  This is a
technique that has been used successfully in another magnetic massive
star, $\theta^1$~Ori~C.  Gagn\'{e} \etal\ (1997, 2005) reported on
multiphase X-ray spectral data from this young O5.5V star.  In that
case the surface field has kiloGauss values (Donati \etal\ 2002).
The field is sufficiently strong in relation to its wind kinetic
energy density that substantial confinement of the wind flow in the
form of an oblique magnetic rotator results, leading to significant
X-ray variability with rotational phase (Babel \& Montmerle 1997a).
Gagn\'{e} \etal\ reported changes in the soft and hard component emission
measures (EM) at about 25\%.

The method has also been used successfully for constraining emission
models in the case of the magnetic early B~star $\beta$~Cep (Favata
\etal\ 2009).  This is a B1~IIIe star that is slowly rotating ($v\sin
i\approx 20$ km s$^{-1}$, Abt \etal\ 2000).  Despite a relatively strong
surface field of about 360~G (Henrichs \etal\ 2000), a {\em Chandra}
and {\em XMM-Newton} study by Favata \etal\ (2009) with pointings at four
different rotational phases of $\beta$~Cep failed to detect the expected
modulation of X-ray emission.  The absence of significant variability
strongly constrains the distribution of magnetically confined hot plasma
in that system.

In the case of $\tau$~Sco, the surface field is slightly stronger
at 500~G than for $\beta$~Cep and considerably weaker than for
$\theta^1$~Ori~C.  Cohen \etal's (2003) {\em Chandra} study of
$\tau$~Sco claimed variability at the 10\% level in the O\,{\sc
viii}\,18.97\AA\ line (0.654\,keV) from an effective exposure of
72~ksec that was distributed over an interval of 120~ksec, or 3\%
of the rotation period of the star (see Fig.~4 of that paper).
$\tau$~Sco was also a regular target for the {International Ultraviolet
Explorer} (IUE), and the P~Cygni lines of N{\sc v}, Si{\sc iv}, and
C{\sc iv} show variable structure with rotational phase.  So, there
is motivation for seeking multiple pointings toward $\tau$~Sco in
an attempt to exploit the ``screening'' effects afforded by stellar
occultation in hopes of ascertaining the location in radius and
azimuth of the hot plasma components in this magnetized massive
star.  A description of the observations and data reduction is
presented in Section~2, followed by an analysis of the multiphase spectral
data in Section~3.  A discussion of our results is given in Section~4 with
concluding remarks given in Section~5.

\begin{deluxetable}{ccccccccccccc}
\rotate
\tabletypesize{\scriptsize}
\tablecaption{SUZAKU Observations of $\tau$\,Sco	\label{tab1}}
\tablewidth{0pt}
\tablehead{ObsID    & HJD--2454696 &  Phase & Exposure &  
\multicolumn{3}{c}{XIS0} & \multicolumn{3}{c}{XIS1} & \multicolumn{3}{c}{XIS3} \\
         &  (days) &  $\phi$ & (ksec) & $\dot{C}_{\rm T}$ & $\dot{C}_{\rm S}$ &  $\dot{C}_{\rm H}$ & $\dot{C}_{\rm T}$ & $\dot{C}_{\rm S}$ &  $\dot{C}_{\rm H}$ & $\dot{C}_{\rm T}$ & $\dot{C}_{\rm S}$ & $\dot{C}_{\rm H}$ }
\startdata
403034050 & 0.31458 & 0.168 &  16.3 & 
$0.806 \pm 0.008$ & $0.306 \pm 0.005$ & $0.500\pm  0.006$ &
$1.554 \pm 0.010$ & $0.859 \pm 0.008$ & $0.695\pm  0.007$ &
$0.827 \pm 0.008$ & $0.324 \pm 0.005$ & $0.503\pm  0.006$ \\
403034060 & 7.19816 & 0.336 & 15.0  & 
$0.808 \pm 0.008$ & $0.317 \pm 0.005$ & $0.491 \pm 0.006$ &
$1.556 \pm 0.011$ & $0.862 \pm 0.008$ & $0.694 \pm 0.007$ &
$0.841 \pm 0.008$ & $0.334 \pm 0.005$ & $0.507 \pm 0.006$ \\
403034010  & 14.04367 & 0.503 & 14.6  & 
$0.825 \pm 0.008$ & $0.329 \pm 0.005$ & $0.496 \pm 0.006$ &
$1.498 \pm 0.011$ & $0.820 \pm 0.008$ & $0.678 \pm 0.008$ &
$0.834 \pm 0.008$ & $0.339 \pm 0.005$ & $0.494 \pm 0.006$ \\
403034020 & 21.07501 & 0.674 & 14.5   & 
$0.829 \pm 0.008$ & $0.325 \pm 0.005$ & $0.504 \pm 0.006$ &
$1.585 \pm 0.011$ & $0.875 \pm 0.008$ & $0.710 \pm 0.008$ &
$0.829 \pm 0.008$ & $0.333 \pm 0.005$ & $0.496 \pm 0.006$ \\
403034040 & 27.42073 & 0.829 & 14.0   & 
$0.810 \pm 0.009$ & $0.326 \pm 0.006$ & $0.484 \pm 0.007$ &
$1.562 \pm 0.012$ & $0.862 \pm 0.009$ & $0.700 \pm 0.008$ &
$0.820 \pm 0.009$ & $0.335 \pm 0.006$ & $0.485 \pm 0.007$ \\
403034030 & 33.41794 & 0.975 & 12.3   & 
$0.849 \pm 0.011$ & $0.338 \pm 0.006$ & $0.511 \pm 0.008$ &
$1.608 \pm 0.015$ & $0.890 \pm 0.011$ & $0.719 \pm 0.010$ &
$0.849 \pm 0.010$ & $0.345 \pm 0.007$ & $0.504 \pm 0.008$ \\ \hline
\enddata
\end{deluxetable} 

\section{Observations and Data Reduction}

The joint Japan/US X-ray astronomy satellite {\em Suzaku} observed
\tsco\ in August and September 2008 at six roughly equally spaced
occasions. The pointings with 12--15\,ksec exposure time 
were spaced approximately 7~days apart to sample one
full rotational cycle of the star which has $P_{\rm rot}=41.033$\,days
(Donati \etal\ 2006b).  Table~\ref{tab1} lists the obsids and dates
of our pointings along with the exposure times in columns 1, 2, and
4.  Other columns in this table will be described later.

{\em Suzaku} carries four X-ray Imaging Spectrometers (``XIS'',
Koyama \etal\ 2007).  The field-of-view (FOV) for the XIS detectors
is $17\arcmin\ \times\,17\arcmin$.  One of the XIS detectors (XIS1)
is back-side illuminated (BI) and the other three (XIS0, XIS2, and
XIS3) are front-side illuminated (FI). The bandpasses are
$\sim$\,0.4\,--\,12\,keV for the FI detectors and $\sim$\,0.2\,--\,12\,keV
for the BI detector.  The BI CCD has higher effective area at low
energies; however its background level across the entire bandpass
is higher as compared to the FI CCDs. The XIS2 detector was not
used in our $\tau$~Sco observations owing to technical difficulties.
Cleaned event files were used to extract the spectra for the science
analysis. Response matrices were generated for each detector and
each pointing using the latest versions of software and the most
recent calibration files.

The angular resolution of the X-ray telescope on-board {\em Suzaku}
is $\approx 2\arcmin$; and so in the XIS imaging, \tsco\ was not
distinctly resolved from the nearby X-ray source 1RXS J163556.0-281149.
According to the recommendations given in the {\em Suzaku} ABC
guide\footnote{http://heasarc.gsfc.nasa.gov/docs/suzaku/analysis/abc/} for
spectral analysis, the spectra were extracted from a $260$ arcsec
region. Inspection of {\em XMM-Newton} images shows that three
serendipitous X-ray sources are present in the spectrum extraction
region. The count-rates of these sources are a few hundreds of times
lower than the count rate of \tsco. No variations in the X-ray
emissions of the two sources located within 2~arcmin of $\tau$~Sco
were reported from previous {\em XMM-Newton} or {\em Chandra}
observations (Mewe \etal\ 2003; Cohen \etal\ 2003).  Therefore, we
believe that the extracted spectra are truly representative of the
X-ray emission of \tsco. However the intrinsic variability of the
serendipitous sources cannot be excluded {\em a priori}.

The background was extracted from an annulus around the source for all
six data sets. The presence of faint point sources in the background
regions cannot be excluded (as seen in the {\em XMM-Newton} images).
However, checks were performed using various background regions selected
in different parts of the {\em Suzaku} FoV, and it was determined that
the influence of the background on the \tsco\ spectrum is negligible.

\begin{figure}[t]
\centering
\includegraphics[width=1.02\columnwidth, angle=0]{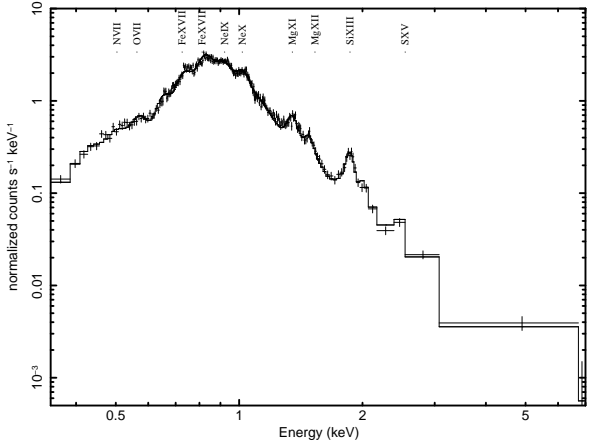}
\caption{The spectrum of $\tau$ Sco obtained by merging the XIS1
spectra obtained on dates HJD 2454696, 2454703, and 2454710 (see
Tab.~\ref{tab1}), along with a best fit model (see text).  The
vertical error bars correspond to $10 \sigma$.  Although no spectral
lines are resolved, a number of typically prominent emission lines
are identified across the top of the figure.
\label{fig1}}
\end{figure}

\section{Analysis}

\subsection{X-ray spectra}

Spectra that were obtained at different phases were found to
be quite similar.  The spectral fitting software {\em xspec} (Arnaud
1996) was used for modeling the data. Based on chi-square fits,
there is little evidence for statistically significant differences
between the six spectra obtained at the different pointings.  Given
the similarity of the spectra, we merged the first three spectral
data sets obtained with XIS1 to achieve higher signal-to-noise.
The resulting spectrum is shown in Figure~\ref{fig1}. The model
spectral fit for this merged set is quite similar to fits obtained
for the spectra from the individual pointing.

A good fit to the spectrum was obtained with a four temperature
thermal plasma model ({\em vapec}) corrected for interstellar
absorption ({\em tbabs}).  Abundances were taken as solar (Asplund
\etal\ 2009), except for C at 0.5 solar, N at 1.4 solar, and oxygen
at 0.7 solar.  These values are similar to the abundances derived
by Morel, Hubrig, \& Briquet (2008) in their study of UV and optical
spectra of $\tau$~Sco.  Also, a ratio N/O of twice solar is in good
agreement with the analysis of {\em XMM-Newton} spectra of \tsco\
by Mewe \etal\ (2003).  In fact, the good agreement between {\em
Suzaku} and {\em XMM-Newton} spectra show that the {\em Suzaku}
data are not seriously affected by the unresolved X-ray sources in
the vicinity of \tsco.  These assumptions lead to fits that have
good statistics and provide a convenient framework for characterizing
and comparing spectra at different rotational phases.

The best fit parameters found from fitting the spectra are shown
in Table~\ref{tab2}.  The hydrogen column density was fixed at a
value of $N_{\rm H}=3\times 10^{20}$\,cm$^{-2}$ from Diplas \&
Savage (1994).  It is interesting to compare the temperature
components from our {\em Suzaku} fits to those of previous studies.
An ASCA spectrum reported by Cohen \etal\ (1997) was fit with a
three temperature component model.  The EM-weighted average temperature of
that fit was $\langle kT \rangle_{\rm ASCA} \approx 1$~keV.  Ours
is much lower at $\langle kT \rangle_{\rm Suz} \approx 0.5$~keV.
The XMM-Newton study by Mewe \etal\ (2003) is $\langle kT \rangle_{\rm
XMM} \approx 0.7$~keV from their RGS data.  However, those authors
find evidence for a hotter component from the EPIC spectrum.  In
their {\em Chandra} study of $\tau$~Sco, Cohen \etal\ (2003) do not
present temperature component fits; however, they do comment that
there is little evidence for hot plasma in excess of 27~MK that was
seen by ASCA.  Following, they note that ASCA has greater sensitivity
at higher energies than does {\em Chandra}.  In their study of
differential emission measures for hot massive stars, Wojdowski \&
Schulz (2005) note that the peak of the EM distribution for $\tau$~Sco
is around 10~MK, or 0.8~keV.

The XIS detectors have even greater
area than did the ASCA-SIS detector, and yet the XIS1 spectrum fails
to show a significant contribution from the S{\sc xv} emission line.
The total EMs derived from the ASCA data and from our {\em Suzaku}
data are actually quite comparable, with values of $\approx~49\times
10^{53}$~cm$^{-3}$ and $\approx~40\times~10^{53}$~cm$^{-3}$, and
yet it appears that the characteristic temperature during our
pointings with {\em Suzaku} is about half as large as when the ASCA
data were obtained.  

\begin{deluxetable}{ccc}
\tablecaption{Spectral Fit Parameters	\label{tab2}}
\tablewidth{0pt}
\tablehead{Component & $kT$ & $EM$ \\
 & (keV) & ($10^{53}$ cm$^{-3}$)}
\startdata
\#1 & $0.11 \pm 0.01$ & $17.0\pm 2.6$ \\
\#2 & $0.34\pm 0.01$ &    $10.4\pm 0.5$ \\
\#3 & $0.71\pm 0.01$ & $\phantom{x}7.2\pm 0.3$ \\
\#4 & $1.52\pm 0.06$ & $\phantom{x}5.2\pm 0.3$ \\
\enddata
\end{deluxetable}

\subsection{Broadband variability}

Our primary objective was to search for evidence of stellar
occultation effects on the X-ray emissions in an attempt to
constrain the location where the soft and hard components are formed.
Since the spectra show no obvious variations between pointings, we
have produced total count rate light curves.  With a relatively
complex distribution of closed and open field lines, we expected
to observe variability with rotational phase.

An ephemeris for the rotation of the field configuration with
Heliocentric Julian date (HJD) is given by Donati \etal\ (2006b) as

\begin{equation}
T({\rm HJD}) = T_0 +P\, E, 
\end{equation}

\noindent where $T_0 ({\rm HJD}) = 2453193.0$, the period $P =
41.033\pm 0.002$ days, and $E$ is the cycle value.  All of our data fell
within the same epoch of rotational cycle 1255.  Rotational phase
values within this cycle number are listed in column~3 of
Table~\ref{tab1}.  Note that Donati \& Landstreet (2009) show more
recent magnetic maps for $\tau$~Sco (see their Fig.~4).  The most
recent of those was taken at about 50 days before the {\em Suzaku}
pointings.  The field has remained stable since the discovery report
of Donati \etal\ (2006b), indicating that any accumulated systematic
errors in rotational phases for the {\em Suzaku} data are less than
2\%.

\begin{figure}[t]
\centering
\includegraphics[width=1.09\columnwidth]{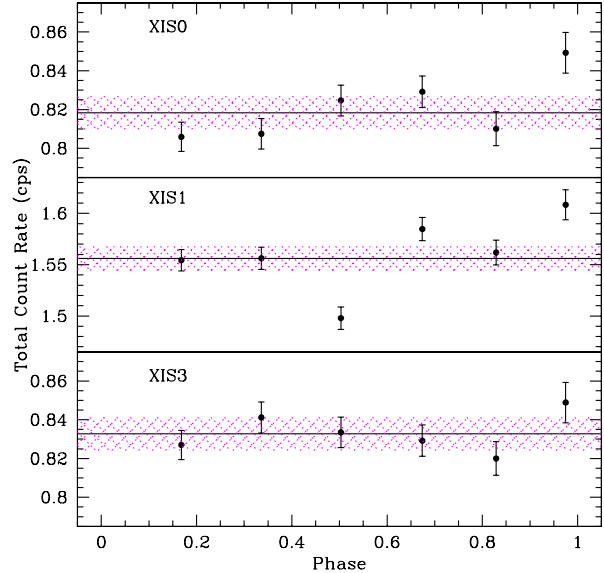}
\caption{Total count rates from Table~\ref{tab1} plotted against 
rotational phase.  The colored hashed region represents
the rms error of the data, as described in the text.
\label{fig2}}
\end{figure}

Spectra for the three detectors with each pointing were summed to
obtain total bandpass count rates, $\dot{C}_T$, in counts per second.
These are listed in the last three columns of
Table~\ref{tab1} and plotted in Figure~\ref{fig2}.
The horizontal lines in each panel represent the mean of the data, as
computed via

\begin{equation}
\dot{C}_{\rm av} = \frac{\sum\,\dot{C}_{\rm i}/\sigma_{\rm i}^2}
	{\sum\,1/\sigma_{\rm i}^2},
\end{equation}

\noindent where $\{\sigma_{\rm i}\}$ are the error values given in
Table~\ref{tab1}.  Note that the error in the mean is given by

\begin{equation}
\sigma_{\rm av} = \frac{1}{6}\,\sqrt{\sum\,\sigma_{\rm i}^2 }
\end{equation}

\noindent for six data points obtained with each detector.
The resultant means and their errors are given in column~2
of Table~\ref{tab3}.

The hashed regions in Figure~\ref{fig2} represent the rms errors
$\sigma_{\rm rms}$ for each data set as given by

\begin{equation}
\sigma_{\rm rms} = \sqrt{\frac{1}{6}\,\sum\,\sigma_{\rm i}^2 }.
\end{equation}

\noindent Given that the errors in the data are all very similar,
the rms error closely represents the values of the individual
measures that determine the respective means. 

\begin{figure}[t]
\centering
\includegraphics[width=0.99\columnwidth]{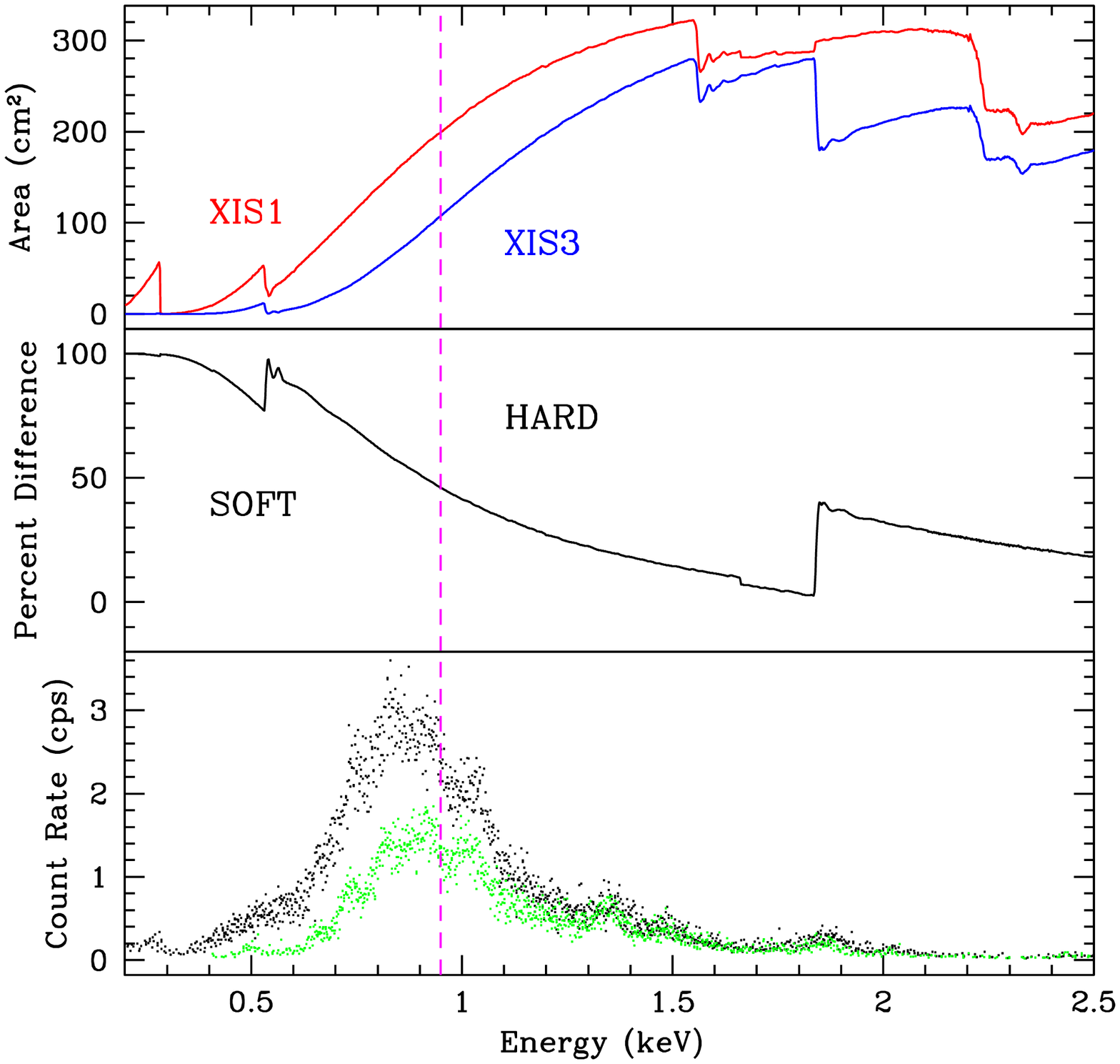}
\caption{A comparison of the areal sensitivities of the XIS0 and
XIS3 detectors with XIS1.  Top shows the XIS1 detector response in
red, whereas blue is for XIS0/XIS3.  A percent difference of these
is plotted in the middle panel.  At bottom is an illustration of
several spectra from XIS1 (higher count rate in black) and XIS3
(lower count rate in green).  The vertical dashed line demarcates
the soft and hard bands used in our analysis.  \label{fig3}}
\end{figure}

Although OB stars do show hot gas components, the hard emission is
not seen with the relative strength at which it appears in $\tau$~Sco
(exceptions being OB~stars that are strongly magnetic or in colliding wind
binary systems).  Thus, it is natural to associate the hot components
seen from $\tau$~Sco with processes involving the star's magnetism.
Then the dominant soft emission most likely arises from nominal
wind shock phenomena as occurs in other OB~winds.  Since an analysis
of He-like ions that form at high temperatures indicate a formation
location of about $1R_\ast$ above the photosphere (Cohen \etal\ 2003),
we therefore anticipated that periodic rotational modulation would affect
the hard X-rays owing to stellar occultation, whereas any variability
of soft X-rays would be mostly stochastic (e.g., Oskinova \etal\ 2001).
Consequently, we have defined ``soft'' and ``hard'' bands as illustrated
in Figure~\ref{fig3}.  The upper panel displays the area sensitivity of
the BI detector (XIS1) as the upper curve in red and of the FI detector as
the lower curve in blue.  The middle panel shows the percent difference
of these.  For reference the lower panel displays a sample of the XIS1
spectra of $\tau$~Sco (brighter points with black dots) and XIS3 spectra
(fainter with green dots).  The vertical dashed line demarcates the soft
and hard bands.  Its placement at 0.95~keV was motivated by the following.
(a) The boundary falls roughly midway along the decline in relative
sensitivity, from the FI chip being about 75\% more sensitive than the
BI chips in the soft band to only 25\% on average in the hard one.
(b) There are roughly equal counts above and below the boundary, so
that count rate errors will be similar.  And (c) the boundary falls at
energies just below several strong emission lines.

\begin{figure}[t]
\centering
\includegraphics[width=0.99\columnwidth]{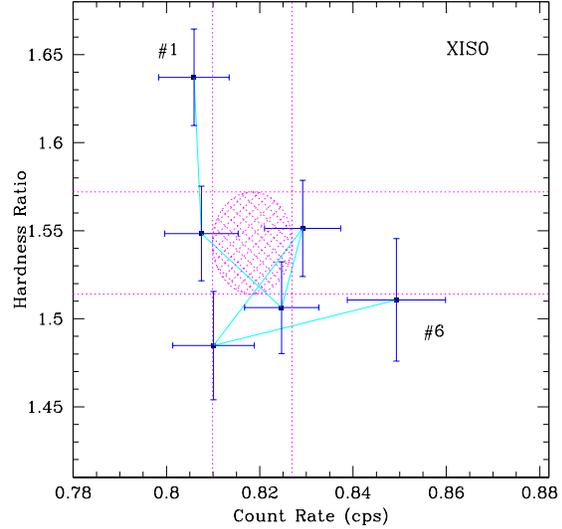}
\caption{A plot of the hardness ratio as
defined in the text plotted against the total count rate
the XIS0 data of $\tau$~Sco.  The hashed region is the rms
error ellipse for these data.  A line connects the data points
in chronological and phase sequence, with \#1 and \#6 marking
the first and last of the pointings (see Tab.~\ref{tab1}).
\label{fig4}}
\end{figure}

\begin{figure}[t]
\centering
\includegraphics[width=0.99\columnwidth]{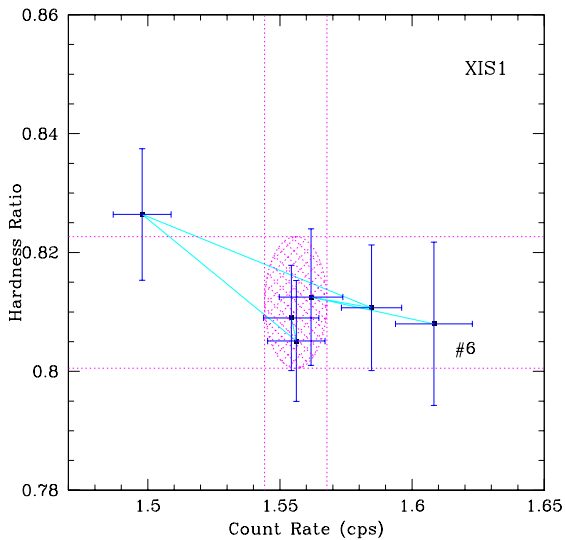}
\caption{As in Fig.~\ref{fig4}, but for XIS1.  Note that
data points are too closely spaced to clearly mark the
first pointing.  \label{fig5}}
\end{figure}

\begin{figure}[t]
\centering
\includegraphics[width=0.99\columnwidth]{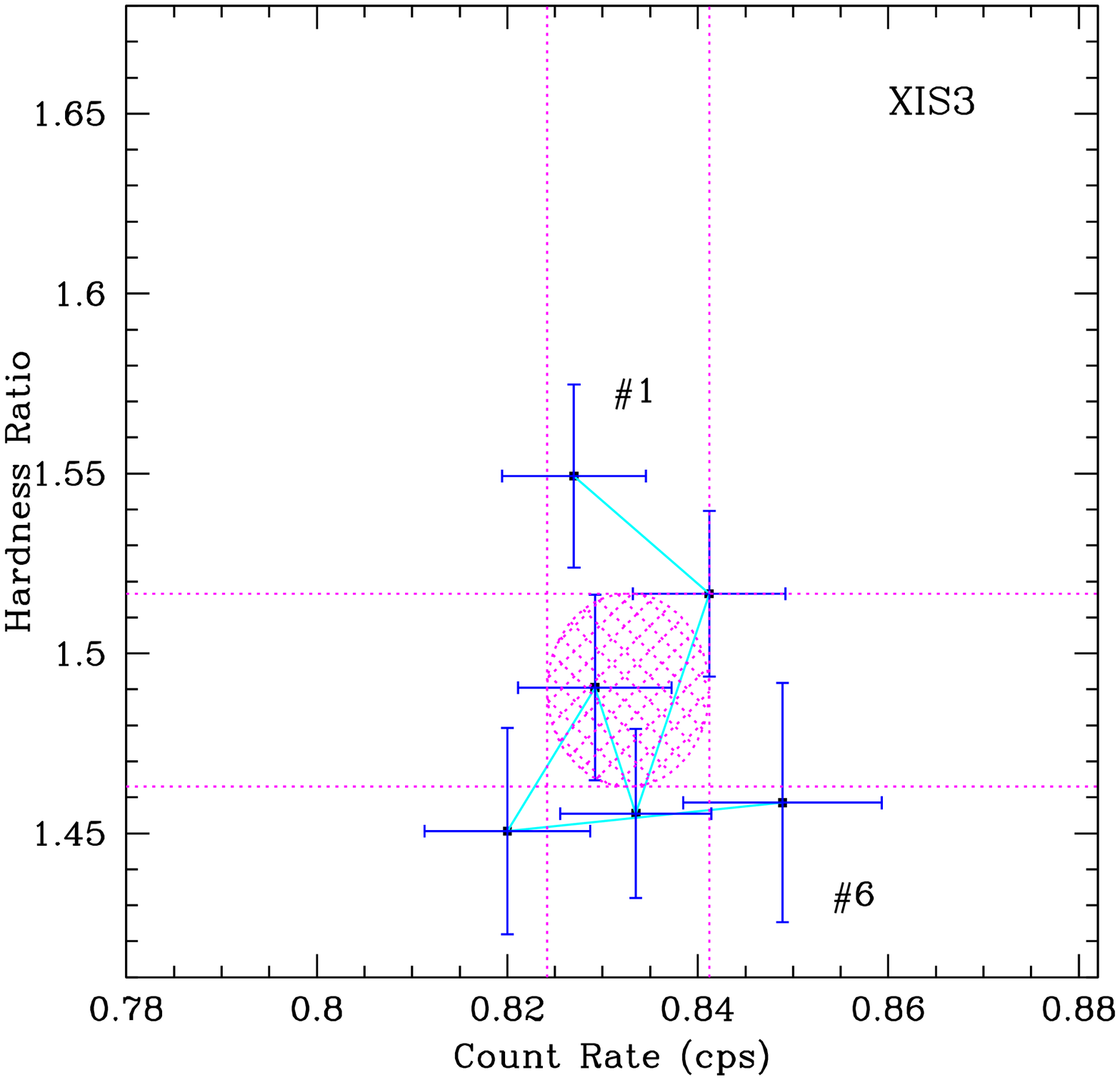}
\caption{As in Fig.~\ref{fig4}, but for XIS3.  
\label{fig6}}
\end{figure}

With these two bands, we introduce the hardness ratio $HR=\dot{C}_H/
\dot{C}_S$ as a ratio of the hard band count rate to that of the
soft band.  Figures~\ref{fig4}--\ref{fig6} plot HR values against
the total count rate for each detector.  An error ellipse is shown
for reference.  This is the rms error ellipse with a center that
is the mean of the HR and total count rate.  Vertical and horizontal
lines that are tangent to the error ellipse are shown as a visual
reference.  A light blue line connects the points in chronological
and phase order, with the first (\#1) and last (\#6) pointings
indicated, except for XIS1 where several points fall too close
together to label clearly which is first.

\begin{deluxetable}{cccccc}
\tablecaption{Statistical Analysis of the X-ray Data	\label{tab3}}
\tablewidth{0pt}
\tablehead{Detector & $\dot{C}_{\rm av}$& $\chi^2_\nu (\dot{C})$ & $HR_{\rm av}$ & $\chi^2_\nu (HR)$ \\ 
 & (cps) & & & }
\startdata
XIS0 & $0.8184 \pm 0.0035$ & 1.82 & 	$1.543 \pm 0.029$ & 1.92 \\
XIS1 & $1.556 \pm 0.005$ & 3.09 & 	$0.812 \pm 0.005$ & 0.69 \\
XIS3 & $0.8327 \pm 0.0035$ & 1.13 & 	$1.490 \pm 0.011$ & 1.53 \\ \hline
\enddata
\end{deluxetable}

Table~\ref{tab3} summarizes the mean values $\dot{C}_{\rm av}$ and
$HR_{\rm av}$ derived from the data and the associated errors in
the means.  Also tabulated are reduced chi-square values $\chi^2_\nu$
with $\nu=5$ for the null hypothesis of no intrinsic source
variability.  Recall that there is a 7.5\% chance of achieving
$\chi^2_\nu = 2$ in the null hypothesis, but only a 1\% probability
of achieving $\chi^2_\nu = 3$.  It thus seems likely that we have
detected variability of the total count rate
in the XIS1 detector, but less so for XIS0 or
XIS3 (however, see the discussion below about the high count rates
for the last phase pointing).  Despite variability in the count
rate, the hardness ratios for XIS1 are remarkably stable.  The
spread in HR values for XIS0 and XIS3 is larger than for XIS1,
although this greater spread is not very statistically significant.

A primary culprit for the variability detected in XIS1 is the third
pointing near phase $\phi\approx 0.5$ as seen in Figure~\ref{fig2}.
Here $\dot{C}$ shows a 3.7\% drop, whereas both XIS0 and XIS3 give
no indication of being much different from the mean.  This could
indicate variability at the softer energies detectable by XIS1,
where XIS0 and XIS3 have less areal sensitivity, especially below
0.5~keV (see Fig.~\ref{fig3}).  The suggestion
is made more plausible by recognizing that in Figure~\ref{fig5},
the lowest count rate pointing of XIS1 corresponds to the highest
HR~value.  Since HR is a ratio of hard to soft, a reduction in the
soft count rate would increase the value of HR.

We have made a comparison of the XIS1 spectrum for the low count
rate point at $\phi \approx 0.5$ against spectra obtained at other
phases.  By binning XIS1 spectra for different pointings and
overplotting them against the low count rate pointing at $\phi=0.50$,
a systematic deficit in the count rate between about 0.8--0.9~keV
is revealed.  Thus the reduced count rate for $\phi=0.50$ along
with a slight increase in HR is not simply a ``drop-out'' in emission
within a narrow energy band, but reflects a somewhat distributed
reduction in emission that lies below 1~keV near the peak of the
spectrum.  Note that we also considered variability in the EMs of
the four temperature components.  In doing so, the four values of
$kT$ were held fixed, as well as the hydrogen column density.
Indeed, the EMs of the two lower temperature components are more
variable than those of the two high temperature ones, as expected.

Interestlying, the {\em ASCA} data that shows evidence of a
considerably higher temperature component than was detected with
{\em Suzaku} occurs at a phase of $\approx 0.45$.  However, the
discrepancy may be a reflection of longer term or stochastic
variability effects unrelated to the rotation period of $\tau$~Sco.

It is also worth noting that all three detectors show rather high count
rates at the last rotational phase.  Deviations from the mean occur at
the level of $3.4\sigma$ for XIS0, $2.2\sigma$ for XIS1, and $2.0\sigma$
for XIS3.  The joint probability for this situation based on each
light curve having independently a normal distribution of data points
is $\approx 10^{-7}$.  Interestingly, Figures~\ref{fig4}--\ref{fig6}
do not indicate that the $HR$ values are particularly anomalous.
With all three detectors showing high count rates by amounts of 3.7\%,
3.3\%, and 2.0\%, respectively, with essentially nominal values of HR,
we suggest that the increased count rate is intrinsic to the source and
essentially ``gray'' in character.  Such a variation might result from
an increase of hot plasma density or volume filling factor.

\section{Discussion}

The critical conclusions from previous X-ray studies are as follows.
The X-ray lines of $\tau$~Sco are strikingly narrow.  The X-ray
spectrum is dominated by a soft component rather typical of other
hot stars.  There is an abnormally strong hard component present as well,
and this hard component in excess of 10~MK has been seen by {\em
ROSAT}, {\em ASCA}, {\em Chandra}, {\em XMM-Newton}, and {\em
Suzaku}.  Moreover, the implied EM for the X-ray emitting plasma
is around $5\times 10^{54}$ cm$^{-3}$.  As pointed out by Cassinelli
\etal\ (1994), this competes with the EM available in the entire
wind.  Of course, it is not possible that the entire wind be at
1--10~MK since UV P~Cygni lines of lower ions are observed.  In
fact, the problem is somewhat exacerbated by the expectation that
in the standard wind shock model of line-driven winds, shocked hot
plasma only begins to be generated at 1.1--1.5 stellar radii into
the wind, implying that even less wind EM is available from which
to generate the observed X-rays.  This problem of large EMs for B
star X-rays was highlighted in Cohen \etal\ (1997).

In terms of generating the hard X-ray emissions, the magnetic field of
$\tau$~Sco is naturally an obvious candidate.  Although the occurrence
of magnetism among OB~stars is not yet known, there are efforts to
determine the incidence rate (e.g., Bagnulo \etal\ 2006; Hubrig
\etal\ 2008; Schnerr \etal\ 2008; Wade \etal\ 2009).  The impact
of magnetism on X-ray emissions is also unclear.  ud-Doula \& Owocki
(2002) used MHD simulations to study the influence of magnetism for
the dynamics of line-driven winds from early-type stars.  Naturally,
the field topology takes on an increasingly dominant role when the
magnetic energy density begins to exceed the ram pressure of the wind
flow.  When the magnetic energy density exceeds the wind kinetic
energy density the field strongly modifies the wind flow leading
to substantial volumes of wind confinement.  This sort of behavior
was explored with a semi-analytic model by Babel \& Montmerle (1997b).


The strong field limit has scored successes in understanding stars
like $\theta^1$~Ori~C and $\sigma$~Ori~E where X-ray data and UV
and optical emission lines can be accurately modeled within a
framework of a magnetically dominated wind flow.  Less clear is how
modest field strengths impact observables.  Although Henrichs,
Schnerr, \& ten Kulve (2005) have reported on correlations between
periodic UV P~Cygni line variability with the incidence of magnetism,
Petit \etal\ (2009) suggest that the X-ray properties of early-type
stars are not so obviously correlated with the presence of surface
magnetism.

Our expectation is that the many unusual features of $\tau$~Sco are
somehow related to the presence of its magnetic field.  Therefore,
the starting point for any analysis should begin with the field
topology.  We possess a model for the magnetic field topology as
extrapolated from the surface magnetograms acquired through
Zeeman-Doppler imaging (Donati \& Collier-Cameron 1997; Donati
\etal\ 1997; Donati \etal\ 1999; Donati \etal\ 2003).  The extrapolation
is performed using the ``Potential Field Source Surface'' method
of Altschuler \& Newkirk (1969) that has been used to extrapolate
the Sun's coronal field from solar magnetograms. A code is used
that was originally developed by van Ballegooijen, Cartledge, \&
Priest (1998). Since the method has been described in Jardine,
Collier-Cameron, \& Donati (2002), only a brief outline of the
method is provided here.

The magnetic field $\bvec{B}$ is written in terms of a flux function
$\Psi$ such that $\bvec{B} = -\bvec{\nabla} \Psi$, so the condition
that the field is potential (i.e., $\bvec{\nabla}\times\bvec{B}
=0$) is satisfied automatically.  The requirement that the field
is divergence-free then reduces to Laplace's equation $\bvec{\nabla}^2
\Psi=0$ with solution in spherical co-ordinates $(r,\vartheta,\varphi)$
given by :

\begin{equation}
\Psi = \sum_{l=1}^{N}\sum_{m=-l}^{l} [a_{\rm lm}r^l + b_{\rm lm}r^{-(l+1)}] \,
         P_{\rm lm}(\theta) \, e^{i m \varphi},
\end{equation}

\noindent where the associated Legendre functions are denoted by
$P_{\rm lm}$.  The coefficients $a_{\rm lm}$ and $b_{\rm lm}$ are
determined by imposing the inferred radial field at the surface
from the Zeeman-Doppler maps and by assuming that at some height
$R_{\rm s}$ above the surface (known as the {\em source surface}),
the gas pressure overcomes the ability of the magnetic field to
confine it. Thus at the source surface, the field lines open up to
become purely radial, and hence  $B_\theta (R_s) = 0$.

\begin{figure*}[t]
\begin{center}
\includegraphics[width=2.in,angle=0]
{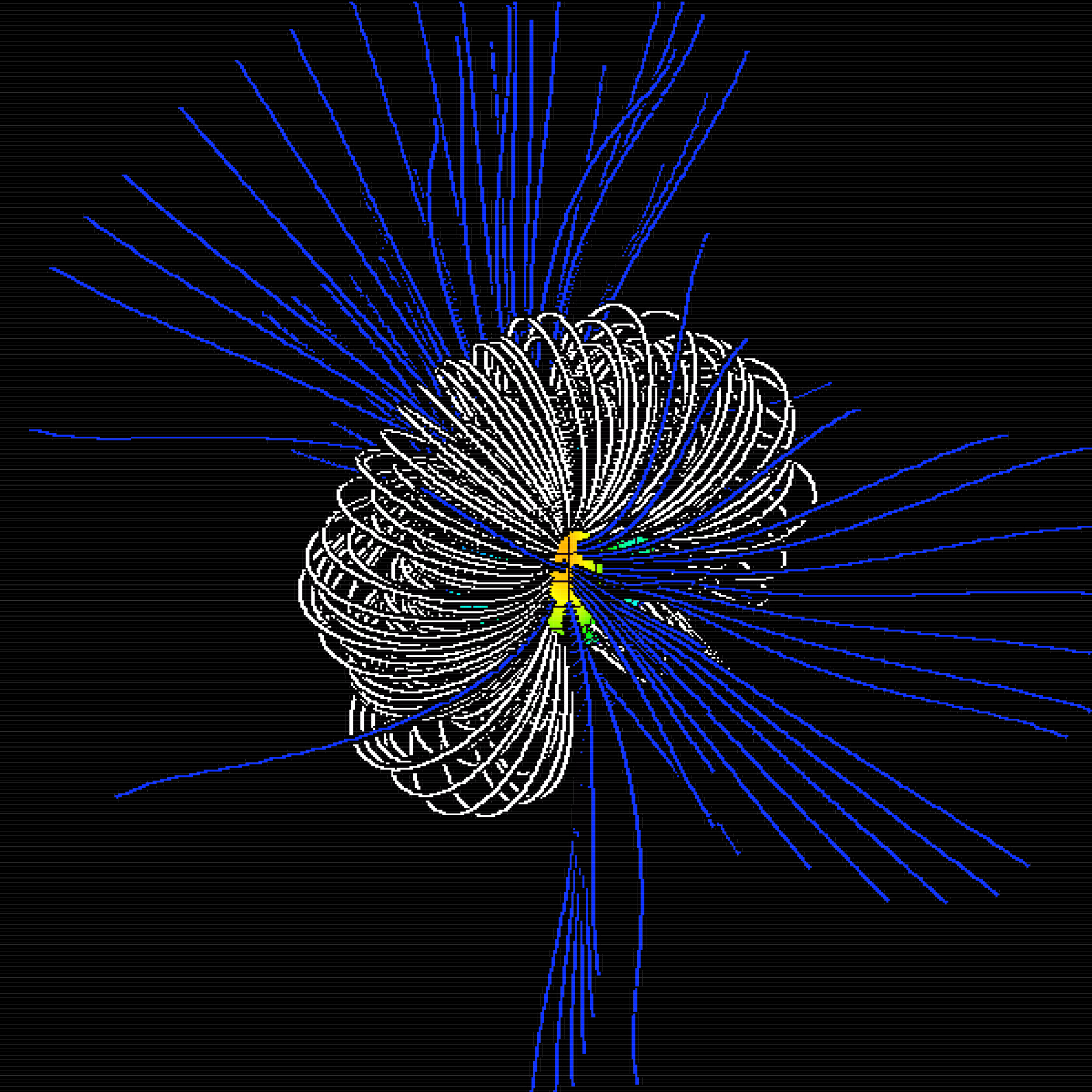}
\includegraphics[width=2.in,angle=0]
{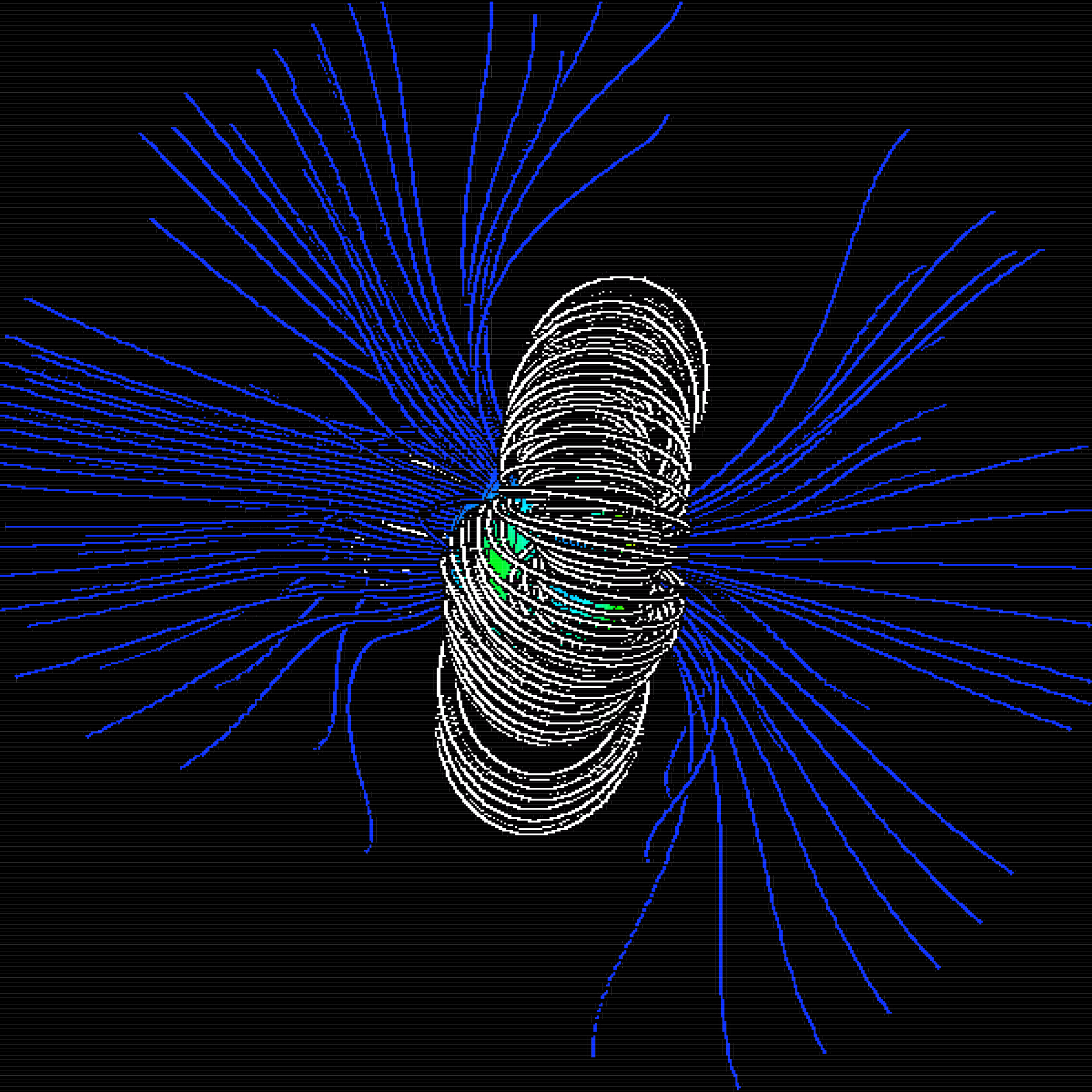}
\includegraphics[width=2.in,angle=0]
{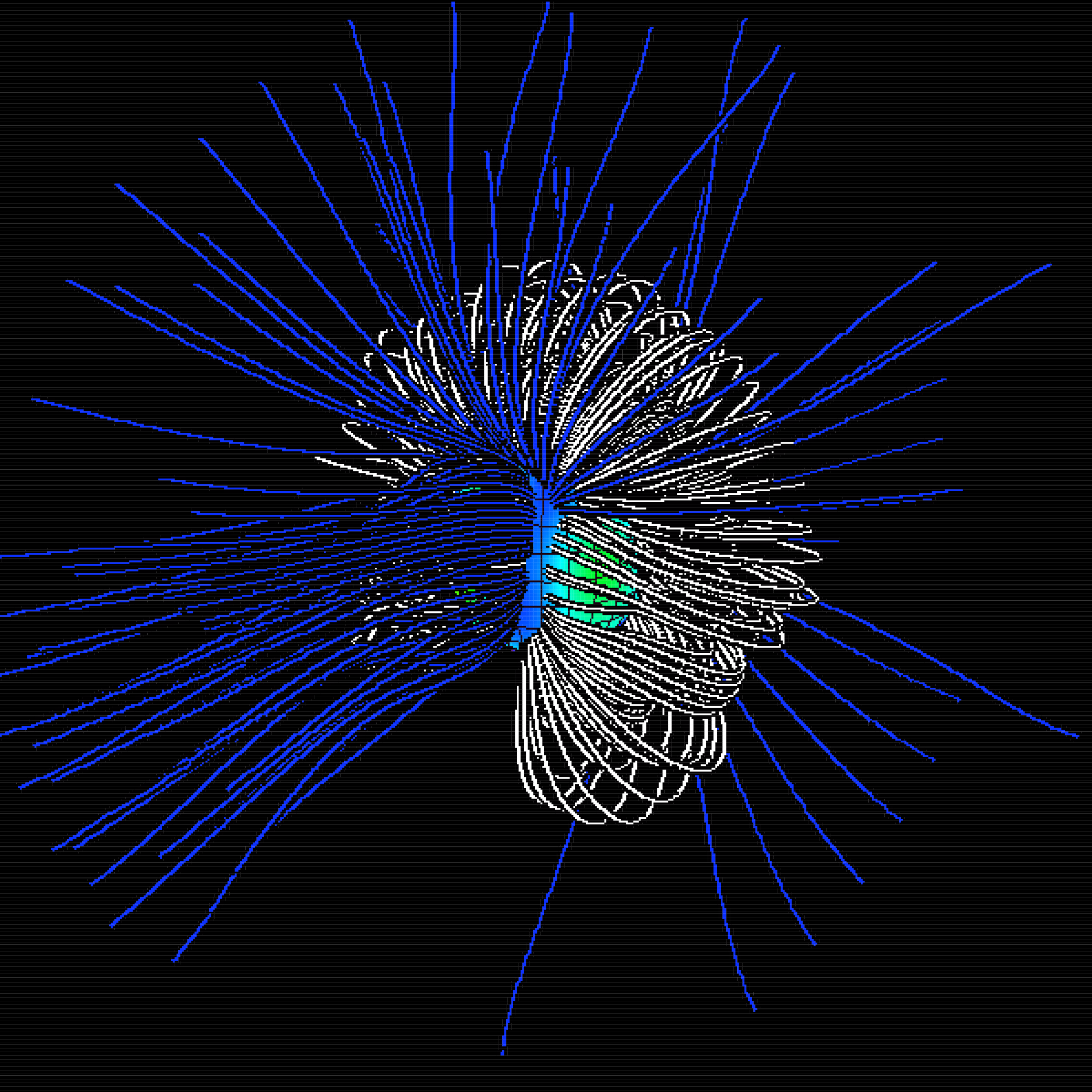}

\includegraphics[width=2.in,angle=0]
{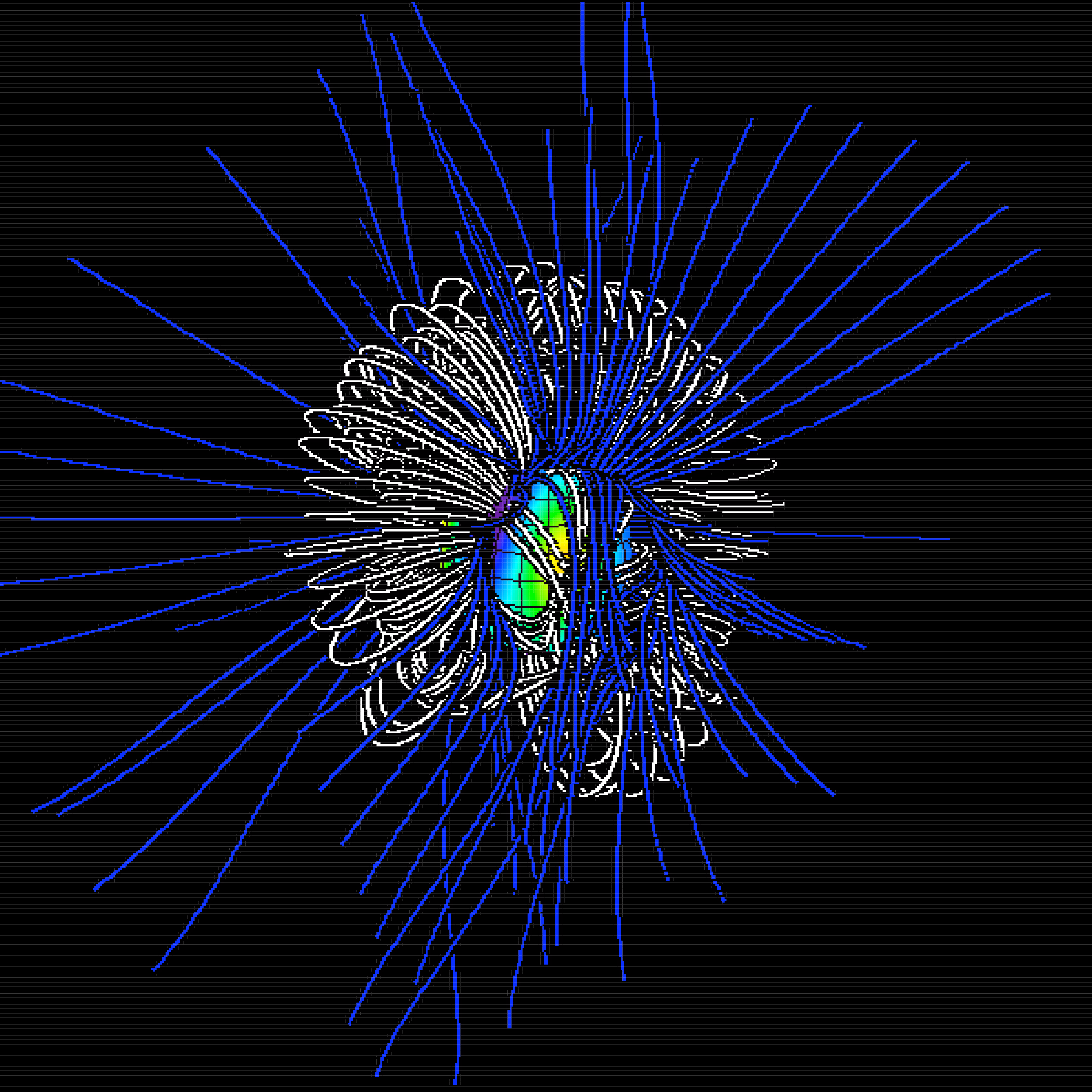}
\includegraphics[width=2.in,angle=0]
{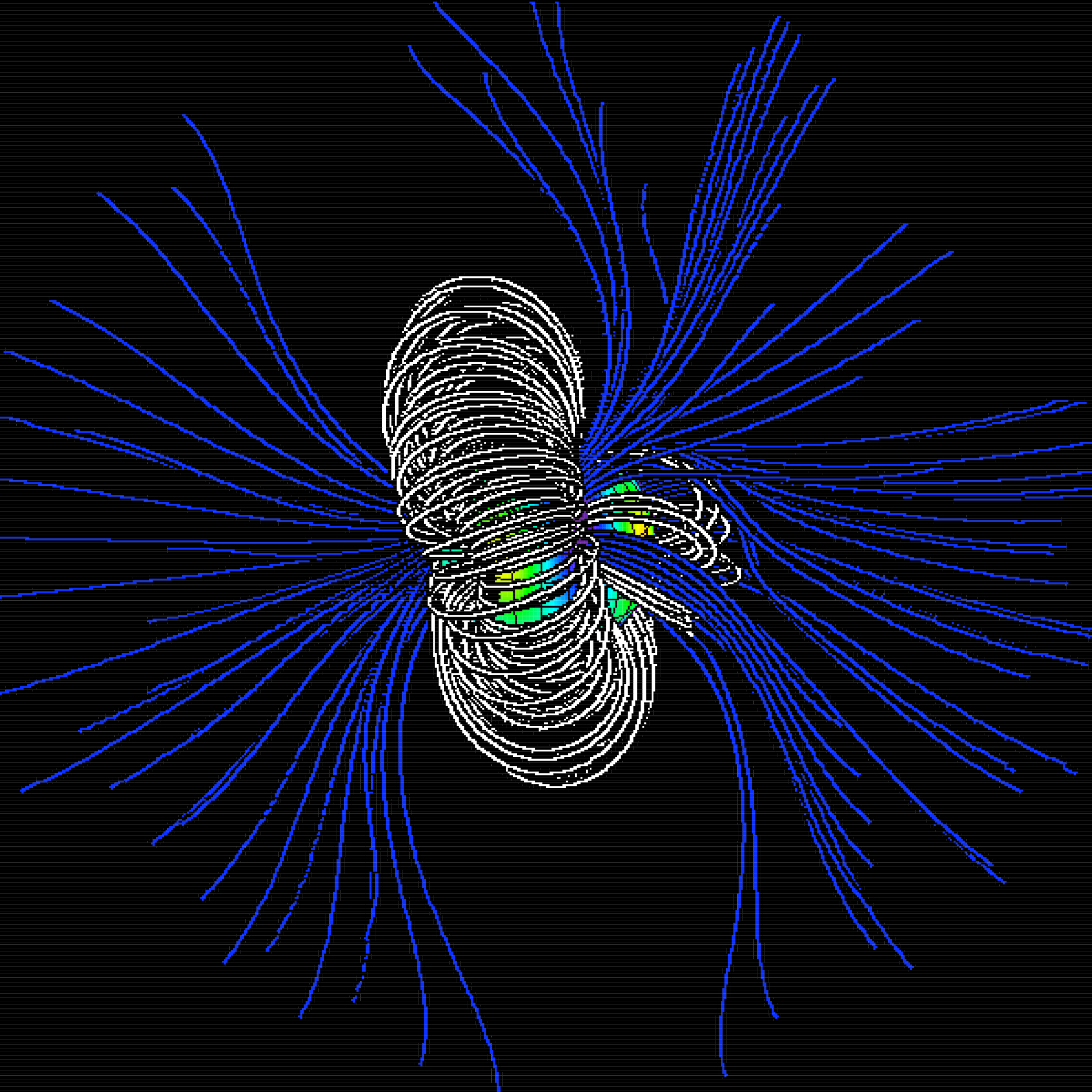}
\includegraphics[width=2.in,angle=0]
{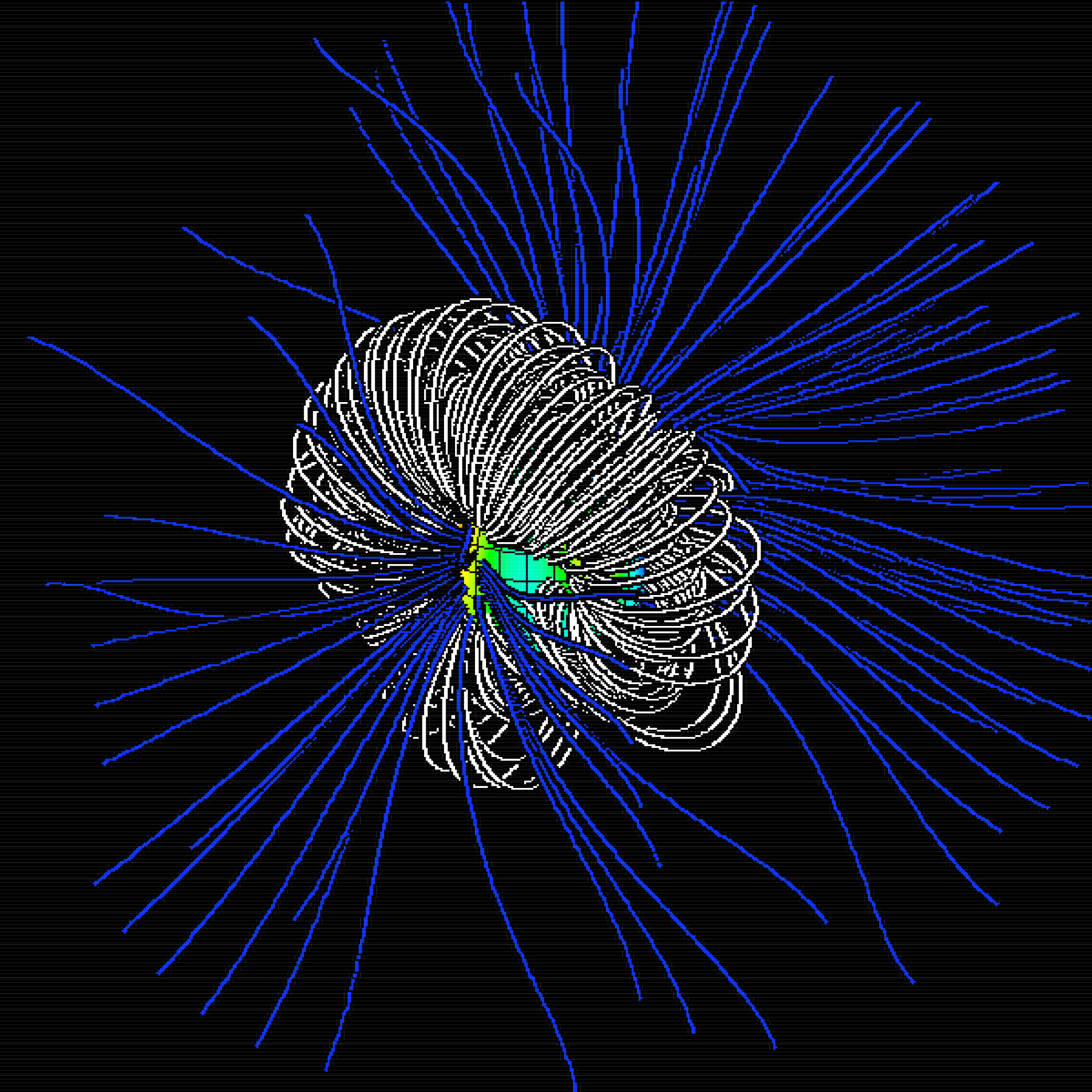}

\caption{Structure of magnetic field lines at rotation phases 0.17,
0.34, and 0.50 (top, left to right) and 0.67, 0.83, and 0.97 (bottom,
left to right).  Closed field lines are shown white, while open
field lines are shown blue.
\label{fig7}}
\end{center}
\end{figure*}

An illustration of the viewable field topology at the rotational
phases of our pointings is shown in Figure~\ref{fig7}.  Along the
top, from left to right, are phases 0.17, 0.34, and 0.50; at bottom
the phases are 0.67, 0.83, and 0.97 from left.  In these figures
the blue curves correspond to open magnetic field lines, and the
white ones are closed loops.  There is a dominant torus-like structure
of closed field lines around an axis that is nearly perpendicular
to the rotation axis.  And on one hemisphere, there is a ``mushroom-like''
bundle of closed fields.  This bundle feature is best perceived off
the right limb for the middle bottom panel at phase $\phi=0.83$.  At
$\phi=0.50$ and 0.67, the ``mushroom'' lies most nearly forefront
of the star, and nearly the entire torus region can be viewed.

In some regions the magnetic field forms closed loops capable of
confining plasma at coronal temperatures, and in other regions the
field lines are open to allow wind flow.  At any given rotation
phase, different field geometries are within view of a distant
observer, leading to the expectation of rotational modulation in
the X-ray emission. If the X-ray emission were to come mainly from
closed loops, then maximum emission should occur when the largest
number of closed loops are within view.  Thus Donati \etal\ (2006b)
predicted two X-ray minima at around phases 0.3 and 0.8. For their
coronal model, the overall magnitude of the EM at any given temperature
depends on two parameters: the extent of the corona (which determines
the magnetic geometry) and the plasma pressure at the coronal base
(which determines the plasma structure). As an example, for a corona
at $2\times 10 ^7$~K and closed loops that extended no further than
1~stellar radius above the surface, they concluded that the closed
loops alone could provide an emission measure of a few times
$10^{54}$~cm$^{-3}$ with a 40\% rotational modulation.

Our {\em Suzaku} XIS pointings reveal a spectrum that is predominantly
soft, with a notable hard component, as has been consistently
observed by other instruments.  For the phases of 0.50 and 0.67,
the ``mushroom'' of closed magnetic loops would be most nearly
forefront, and yet somewhat surprisingly, no increase in hard X-rays
was detected.  Instead, we find that the XIS0 and XIS3 show little
evidence for variability, but that the XIS1 detector with about
twice the sensitivity of the other detectors has a {\em drop} in
total count rate near phase $\phi\approx 0.5$, with a $3.7\sigma$
deviation from the mean.  From a comparison between XIS1 spectra at
different pointings, there appears to be a slight deficit in soft
emission at $\phi\approx 0.5$ at energies below 1~keV as compared
to other pointings.  This is suggested in Figure~\ref{fig5} where the
lowest count rate for XIS1 also has the highest HR ratio.  We also
note that all three detectors show an essentially ``gray'' increase
in count rate at phase 0.98 by about 3\%, which corresponds to a
time when the ``mushroom'' bundle of closed field lines 
is occulted by the star.

Why is the hard emission not more variable?  The HR values for XIS1
are amazingly stable, and although HR values show greater dispersion
for XIS0 and XIS3 detectors than for XIS1, their variability is not
statistically especially significant.  Our expectation was that the
hard emission would form in the vicinity of the photospheric magnetic
field, which asymmetrically distributed about this slowly rotating
star.  We therefore predicted that a rotational modulation could
arise from stellar occultation.  If the bulk of the hard emission
were to form at a characteristic radius $r_{\rm hard}$, then the
geometric dilution factor

\begin{equation}
W(r) = \frac{1}{2}\,\left(\, 1-\sqrt{1-R_\ast^2/r^2}\,\right)
\end{equation}

\noindent represents the areal fraction of a spherical shell of
that radius that is occulted by the star, thus $\Delta \Omega_{\rm
occ}/4\pi = W(r_{\rm hard})$.  If the hottest plasma forms
characteristically at $r_{\rm hard} \approx 2R_\ast$ as indicated
by the {\em Chandra} analysis of Cohen \etal\ (2003), then a fraction
$W \approx 7\%$ of the source region would be occulted.

If all of the X-rays were associated with the magnetic field of $\tau$~Sco
and generated at $2R_\ast$, then given the distributed nature of the
field reconstruction as shown in Figure~\ref{fig7}, one might expect
$W$ to represent a characteristic scale of X-ray variability on the
rotation phase.  But this is a vastly oversimplified picture.  In reality,
$\tau$~Sco has a complex field topology that is relatively strong which can
alternately confine wind plasma in closed loops and modify wind flow in
sectors of open field lines.  There are a variety of emissive components
at varying heights above the photosphere as a function of latitude and
azimuth about the star.  It is the recognition of these complexities that
motivate the detailed field reconstruction presented in Jardine \etal\
(2002) and that was briefly summarized above.

With its higher sensitivity, we interpret the modest variability
in the total count rate of XIS1 with relatively little change in
the hard spectrum as a {\em moment} constraint on the complex
integral relations involving geometry and emissive components that
determine the observed emission from an unresolved source.  The
overall relative absence of variability with phase should provide
useful constraints for future modeling efforts that take into account
both the wind flow on open field lines and confined plasma in closed
field loops.

In addition, a recent development is the identification of $\tau$~Sco
analogues.  For a long time, the combination of unusual properties
displayed by $\tau$~Sco had been unique.  This may no longer be the
case.  The early B~stars HD~66665 and HD~63425 had been noted to
show unusual UV lines similar to $\tau$~Sco (Massa, priv comm).
Recently, magnetic fields have been detected in these two stars by
the MiMeS collaboration\footnote{www.physics.queensu.ca/~wade/mimes}
(Petit \etal, in prep).  It would be useful to learn if these two
stars are also hard X-ray emitters.  If so, then $\tau$~Sco would
no longer be a class unto itself, and the key to understanding
$\tau$~Sco may derive from studies of this set of three stars.

\section{Conclusions}

In Donati \etal's (2006b) discovery paper of $\tau$~Sco's magnetism,
the authors predicted X-ray modulations on the rotation period of the
star could be as much as $\sim 40\%$.  In particular, two main eclipses
were expected at phases of around 0.3 and 0.8 in conjunction with observed
enhanced UV line absorptions at those times.  Our pointings include phases
close to those targets, but no variations anywhere near that amplitude
were detected.  It is possible that strong variability could have been
missed if it had occurred in a relatively short interval of phase.
However, rapid and high amplitude variability would appear to run
counter to expectations based on the UV data.  Donati \etal\ noted that
an absence of the predicted high amplitude X-ray modulation could be an
indicator of smaller scale magnetic loops and confined hot gas across the
stellar surface.  Such loops would have evaded detection in their study.

The most unexpected result of our study is that there is greater
variability of the soft X-ray component than of the hard one.
We nominally expected the soft emission to be associated with wind
shocks similar to OB stars, and the unusually bright hard emission
to be associated with the star's magnetic field.  With closed
loops extending out to a few stellar radii, variability of the hard
emission was supposed to occur as a result of stellar occultation of
the rotating magnetosphere; instead, the more sensitive XIS1 detector
shows a statistically significant low count rate event because the soft
component drops more than the hard one.  This came as a surprise.

It thus appears that more effort will be needed to clarify the
nature of $\tau$~Sco's hard emission.  Although the observed soft
component is relatively nominal for a massive star wind of $\tau$~Sco's
early B spectral type, the standard wind shock theory most certainly
does not predict large emission measures well in excess of 10~MK
(e.g., Lucy 1982; Owocki, Castor, \& Rybicki 1988; Feldmeier, Puls,
\& Pauldrach 1997).  Our {\em Suzaku} study has not, in itself,
resulted in a definitive explanation of how (or where) $\tau$~Sco's
hard emission is produced.  There are a number of possible models
that could be employed to interpret the new {\em Suzaku} data in
conjunction with data from other wave bands, such as the unusual
UV P~Cygni line shapes.  We suggest that additional multiwavelength
monitoring will be crucial for formulating a full picture of the
circumstellar environment of $\tau$~Sco.

In particular, the phase coverage of our {\em Suzaku} study is still
sparse.  Plus it would be valuable to have multi-epoch data to
determine whether the high count rates near phase 0.98 persists.
In addition, the presence or absence of linear polarization would
help to define the wind structure and/or the photospheric brightness
distribution at inner wind radii.  At IR wavelengths Brackett lines
have been observed to be in emission whereas H$\alpha$ is in
absorption for $\tau$~Sco (Waters \etal\ 1993).  Although originally
interpreted as a low density disk, Zaal \etal\ (1999) have suggested
that the Brackett line emission could result from NLTE effects in
the atmosphere of $\tau$~Sco.  A study of variability in these lines
would help to determine their origin and place additional constraints
on the density and temperature stratification that could be used
in conjunction with the Zeeman maps and the X-ray and UV data to
produce a more coherent picture of the $\tau$~Sco environment.

\begin{acknowledgements}

We gratefully acknowledge an anonymous referee who made comments that
have improved this paper.  We thank the Suzaku team for performing this
observation and providing software and calibration for the data analysis.
Thanks also to the Suzaku helpdesk, especially K. Hamaguchi, for their
assistance.  This research has made use of NASA's Astrophysics Data
System Service and the SIMBAD database, operated at CDS, Strasbourg.
This research was supported in part by NASA grant NNX09AH24G (RI) and
DLR grant 50OR0804 (LMO).

\end{acknowledgements}

\end{document}